\documentclass[onecolumn,preprintnumbers,amsmath,amssymb,superscriptaddress]{revtex4}
\usepackage{graphicx,subfigure,epsfig,epstopdf}
\usepackage{graphicx}% Include figure files
\usepackage{dcolumn}% Align table columns on decimal point
\usepackage{bm}% bold math
\usepackage{amsmath,amssymb,upgreek}
\input{epsf}
\input{epsfx}
\makeatletter

\newcommand{\Rmnum}[1]{\expandafter\@slowromancap\romannumeral #1@}
\makeatother

\begin{document}
\title{High performance graphene oxide-based humidity sensor integrated on a photonic crystal cavity}

\date{Jan. 10, 2017}

\author{Xuetao Gan}
\email{xuetaogan@nwpu.edu.cn}
\affiliation{MOE Key Laboratory of Space Applied Physics and Chemistry, and Shaanxi Key Laboratory of Optical Information Technology, School of Science, Northwestern Polytechnical University, XiÕan 710129, China}
\author{Chenyang Zhao}
\affiliation{MOE Key Laboratory of Space Applied Physics and Chemistry, and Shaanxi Key Laboratory of Optical Information Technology, School of Science, Northwestern Polytechnical University, XiÕan 710129, China}
\author{Qingchen Yuan}
\affiliation{MOE Key Laboratory of Space Applied Physics and Chemistry, and Shaanxi Key Laboratory of Optical Information Technology, School of Science, Northwestern Polytechnical University, XiÕan 710129, China}
\author{Liang Fang}
\affiliation{MOE Key Laboratory of Space Applied Physics and Chemistry, and Shaanxi Key Laboratory of Optical Information Technology, School of Science, Northwestern Polytechnical University, XiÕan 710129, China}
\author{Yongjiang Li}
\affiliation{MOE Key Laboratory of Space Applied Physics and Chemistry, School of Science, Northwestern Polytechnical University, XiÕan 710129, China}
\author{Jianbo Yin}
\affiliation{MOE Key Laboratory of Space Applied Physics and Chemistry, School of Science, Northwestern Polytechnical University, XiÕan 710129, China}
\author{Xiaoyan Ma}
\affiliation{MOE Key Laboratory of Space Applied Physics and Chemistry, School of Science, Northwestern Polytechnical University, XiÕan 710129, China}
\author{Jianlin Zhao}
\email{jlzhao@nwpu.edu.cn}
\affiliation{MOE Key Laboratory of Space Applied Physics and Chemistry, and Shaanxi Key Laboratory of Optical Information Technology, School of Science, Northwestern Polytechnical University, XiÕan 710129, China}

%\section{Introduction}
\begin{abstract}
We report a high performance relative humidity (RH) microsensor by coating a few-layer graphene oxide (GO) flake over a photonic crystal (PC) cavity. Since the GO layer has a high water-activity and interacts with the evanescent cavity mode strongly,  the exposure of GO-PC cavity in varied humidity levels results in significant resonant wavelength shifts, showing a slope of 0.68 nm$/\%$RH in the range of 60\%$\sim$85\% RH. By interrogating the power variation of the cavity reflection, the microsensor presents an ultrahigh sensitivity exceeding 3.9 dB$/\%$RH. Relying on the unimpeded permeation of water molecules through the GO interlayers and microscale distribution of the cavity mode, the integrated sensor has a response time less than 100 ms, which promises successful measurements of human breathing.  Combining with the ease of fabrication, this high performance RH sensor has potential applications requiring optical access, device compactness, and fast dynamic response.  
\end{abstract}
%\pacs{42.50.Ct, 42.50.Dv, 42.70.Qs, 78.67.Hc}
\maketitle

%\section{Introduction}

Detection of relative humidity (RH) is extremely important in environment monitoring, medicine and indoor air quality control, and chemical industry processing. Two-dimensional (2D) materials, including graphene, transition metal dichalcogenides, and phosphorene, have shown great promises for RH sensing due to their high surface-to-volume ratio, low noise and sensitivity of electronic properties to the surrounding variations~\cite{Smith2015,Yasaei2015,Transistors2013}. Here, we report high performance 2D material-based RH sensing could be optically accessed  as well by integrating the active material with a photonic crystal (PC) cavity. The PC cavities are fabricated in a silicon on insulator (SOI) wafer, facilitating a massive production in the CMOS industry. Because of the high refractive index of silicon, the fabricated PC cavities have resonant modes strongly confined in a cubic-wavelength mode volume with high quality ($Q$) factor ($\sim{10^3}$). By covering mono- or few-layer 2D materials onto the silicon PC cavity, the evanescent field of the resonant mode would interact with the 2D materials strongly, as demonstrated in experiments of nonlinear optical processes~\cite{Shi2015,TingyiGu}, and high-performance optoelectronic devices~\cite{Gan2013b,Gan2014,Gao2015,Shiue2013}. This strong light-matter interaction makes the resonant wavelength of the PC cavity very sensitive to the refractive index change of the 2D materials, even only for a single atomic layer~\cite{Gan2013a}. Therefore, if a humidity sensitive 2D material is integrated with a PC cavity, the RH levels could be monitored precisely by measuring the wavelength or intensity variations of the resonant peak. 

\begin{figure}[th!]\centering
\includegraphics[width=6in]{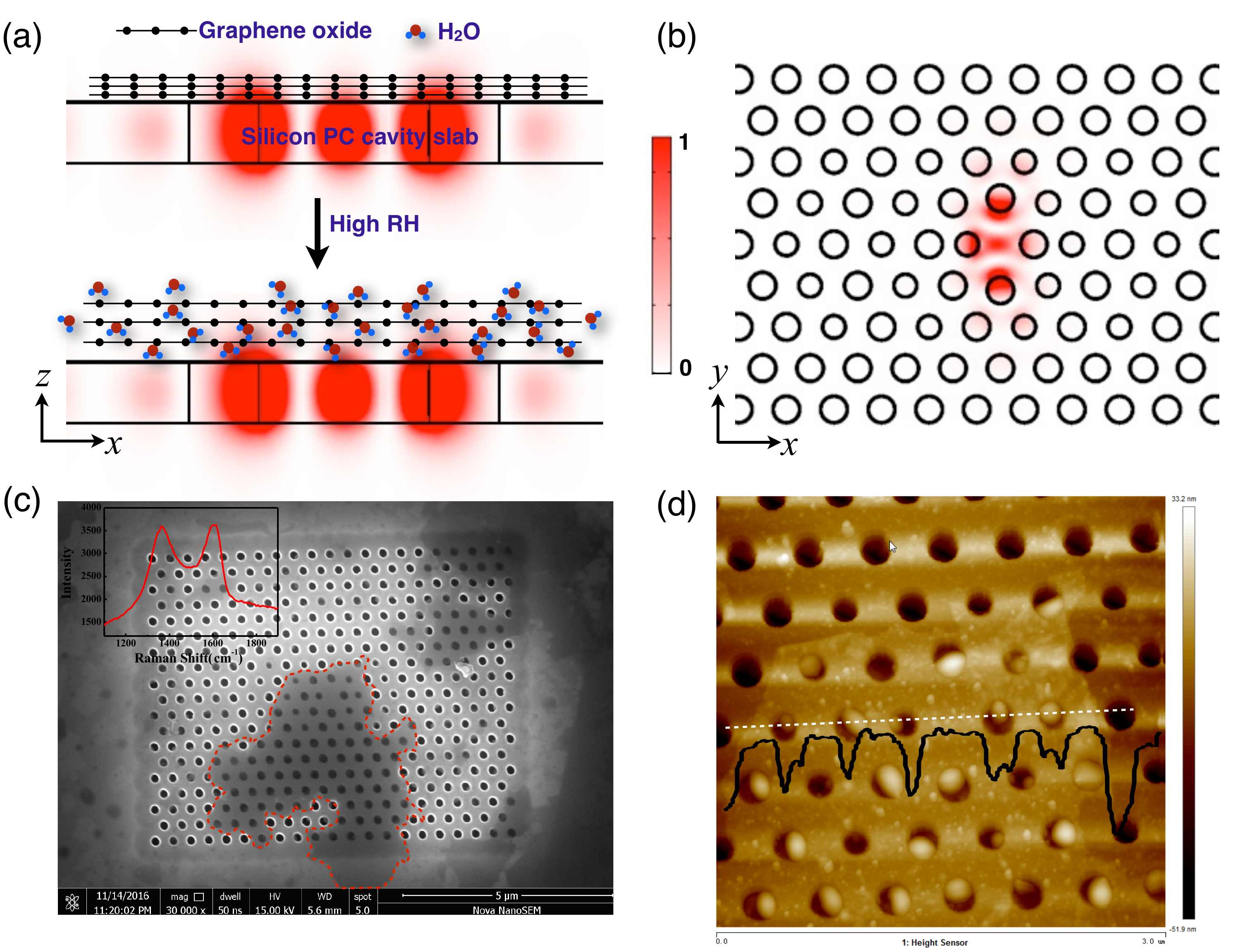}
\caption{{\small (Color online) (a)Schematic of the interaction between GO flake and resonant mode of the PC cavity, where the GO layer is expanded under a high RH because water molecules permeate into the GO interlayer. The red regions indicate the resonant mode distribution cross the silicon slab. (b) Simulated energy field distribution of the resonant mode with a microscale dimension. (c) Scanning electron microscope image of the GO-PC cavity device, where the GO flake is indicated by the dashed line. The inset displays the Raman spectrum of the GO flake. (d) Atomic force microscope image of the GO-PC cavity device, showing a GO layer thickness of 9 nm. }}
\label{fig:sche} 
\end{figure}

In our experiment, we choose few-layer graphene oxide (GO) as the humidity active material. While chemical vapor deposition or mechanical exfoliated graphene provide high layer quality, GO has hydroxyl, carbonyl, and carboxyl functional groups to absorb moisture easily, and it could be massively produced and low-cost. Already, GO films have been integrated onto optical fiber devices for optical humidity sensings, presenting wide dynamic range, high sensitivity, and fast response~\cite{Xiao2014,Wang2016}. Here, the PC cavity is employed to achieve much stronger interaction between GO layers and optical fileds as well as compactly chip-integrated device footprint, which promises improved sensing performances. As the RH level increases (decreases), water molecules are adsorbed into  (desorbed out of) the GO layers, which expands (shrinks) thickness of the GO layer on the PC cavity, as schematically shown in Fig.~\ref{fig:sche}(a). The strong light-matter interaction allows the thickness variation of the GO flake to shift  the resonant wavelength significantly~\cite{Gan2012}. We observe the resonant wavelength of the GO-PC cavity shifts with a slope of 0.68 nm$/\%$RH under the humidity range of 60\%$\sim$85\%RH. In addition, because PC cavity has a wavelength-scale mode distribution,  the size of the employed GO flake could be designed into a microscale, which enables fast interlayer adsorption and desorption of water molecules. In a fabricated GO-PC cavity sensor, a fast response time less than 100 ms is achieved, promising successful responses of human breathing. 

The employed PC cavities are fabricated on a SOI wafer with a 220 nm thick top silicon layer. The PC patterns are defined by electron beam lithography (EBL) on a electron beam resist (ZEP 520), which is spin-coated on the wafer. After the resist development,  the EBL patterns are transferred into the top silicon layer with an inductively coupled plasma (ICP) dry etching. After removing the electron beam resist, the bottom oxide layer is wet etched with a diluted hydrofluoric acid to suspend the PC cavities. The PC cavity is designed with a point-shifted defect by shifting the vertical and horizontal holes around the cavity center, as indicated in Fig.~\ref{fig:sche}(b). To assist the vertical coupling of the cavity mode, air-holes around the cavity defect are shrunk to break the mode symmetry~\cite{Narimatsu}. The lattice constant of the PC pattern is designed as 450 nm to obtain modes resonant around 1500$\sim$1600 nm. The energy field distribution (in the $x-y$ plane) of the resonant mode is also plotted by superimposing over the PC pattern in Fig.~\ref{fig:sche}(b), which is simulated using a three-dimensional finite element method (COMSOL Multiphysics). Relying on the high refractive index of the silicon slab and the large photonic bandgap, the resonant mode is strongly confined around the defect region with a $Q$ factor of 9,000 and a mode volume of 0.28$(\lambda/n)^3$. The energy field distribution (in the $x-z$ plane) over the slab cross-section is displayed in Fig.~\ref{fig:sche}(a), showing the evanescent interaction between the resonant mode and the GO flake.  

The GO is fabricated with Hummer's method, which is dispersed in water by sonication to make a stable suspension of GO crystallites. For the sensor device fabrication, GO flakes with thicknesses of few nanometers are chosen with the assistance of centrifugation. The GO solution is drop-casted onto the silicon PC chip. After the water vaporization, separated flakes are obtained due to the low concentration of GO flakes.  Figure~\ref{fig:sche}(c) shows the scanning electron microscope image of the fabricated device, where the dashed line indicates a GO flake coated over the defect region of the PC cavity. By optically pumping the GO flake with a 532 nm continuous-wave laser, the Raman spectrum is acquired with two resonant peaks around 1350 cm$^{-1}$ and 1600 cm$^{-1}$, as shown in the inset. The morphology of the GO coated PC cavity is examined by an atomic force microscopy (AFM), as displayed in Fig.~\ref{fig:sche}(d).  The integrated GO flake has a uniform thickness of $\sim$9 nm, corresponding to a layer number of 15 by assuming an interlayer distance of $\sim$0.6 nm. The film heights over the PC air-holes indicate the few-layer GO flake is freestanding without breaking even after the AFM measurement, and the GO film contacts with the silicon substrate conformally. Comparing the PC structures shown in Figs.~\ref{fig:sche}(b) and~\ref{fig:sche}(d), the GO flake overlaps with the cavity region and could interact with the resonant mode effectively. 

\vspace{12pt}

To implement the RH sensing, the GO-PC cavity is loaded in a home-made chamber to maintain a stable RH level, which has a glass window to allow the access of optical signal. Different humidity levels in the chamber are achieved by mixing a dry air and a saturated humid air with the controlled air-flows~\cite{Zhao2016}. A commercial RH sensor is inserted into the chamber as well to monitor the real-time RH. The chamber is mounted onto a thermo-electrically controlled stage to guarantee the constant temperature during the RH sensing. The GO-PC cavity is optically characterized using a cross-polarized microscope setup. With an external excitation light source, the cavity reflection is acquired to examine the resonant modes. The cavity excitation and reflection are coupled vertically using an 50$\times$ objective lens with a numerical aperture of 0.42. With the orthogonally polarized input and output, the reflection of the cavity mode could be distinguished with a high signal to noise ratio~\cite{Gan2013c}. The employed light source is a narrowband tunable telecom laser, and the cavity reflection is detected by a photodiode. By tuning the laser wavelength with a step of 0.005 nm, the cavity's reflection spectra are obtained. 

\begin{figure}[th!]\centering
\includegraphics[width=6in]{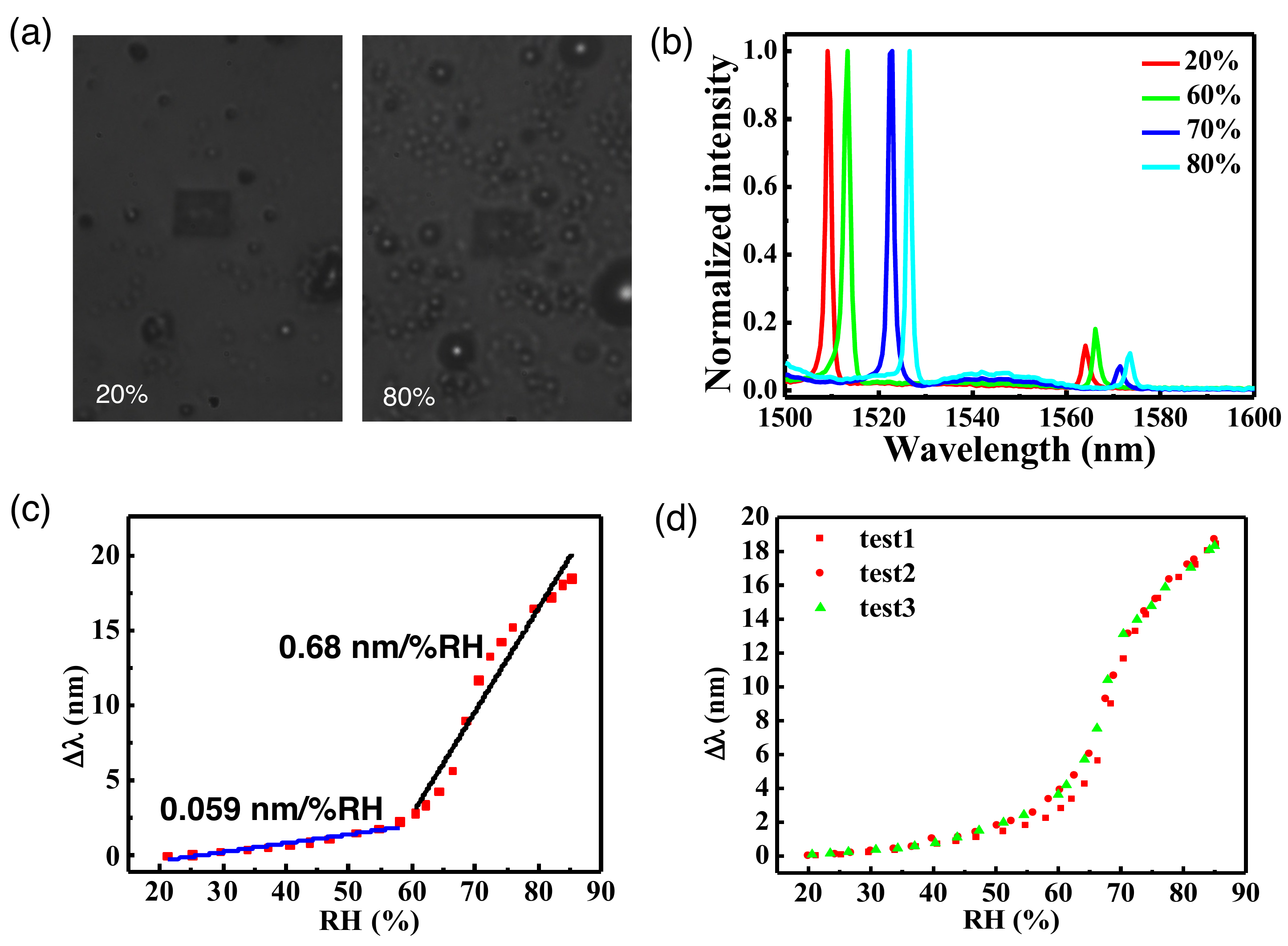}
\caption{{\small (Color online) (a) Optical images of the device under 20\%RH and 80\%RH, which are captured by the camera of the experimental setup.  (b) Reflection spectra of the GO-PC cavity with different RH levels. (c) Resonant wavelength shift versus the humidity levels during a range of 20\% and 85\%RH, where the lines are the linear fittings. (d) Repeatability test of the RH sensing. }}
\label{fig:spe} 
\end{figure}

Figure~\ref{fig:spe}(a)  displays optical images of the GO-PC cavity under 20\% RH (left) and 80\% RH (right), which are captured by the camera of the experimental setup. Condensation water clearly exhibits over the separated GO flakes with high humidity levels, indicating their effective water adsorption. The cavity reflection spectra are acquired as well with different RH levels, as shown in Fig.~\ref{fig:spe}(b).  Two resonant modes are observed over the wavelength range of 1500$\sim$1600 nm, and the one at the shorter wavelength corresponds to the simulated mode shown in Fig.~\ref{fig:sche}(b). As the RH increases, both modes red-shift gradually, while the reflection powers and $Q$ factors undergo nearly no variations. We record the resonant wavelength variations of the mode at shorter wavelength by changing the RH continuously in a range between 20\% and 85\%, which is the limited RH level achieved by our experimental facility. The results are plotted in Fig.~\ref{fig:spe}(c). Different from the performances of optical RH sensors enabled by other humidity active materials~\cite{Zhao2016}, the GO-based RH sensor presents a strongly RH-dependent behavior~\cite{Xiao2014}. For the humidity level smaller than  $60\%$RH, the resonant wavelength varies slowly with respect to the increased RH, showing a linear slope of  $\sim$0.059 nm/\%RH. The sensitivity increases to a remarkably high value of 0.68 nm/\%RH for the RH higher than $60\%$, as indicated by the fitting lines in Fig.~\ref{fig:spe}(c). While the wavelength sensitivity in the range of $20\%\sim60\%$RH  is not as high as that in the range of $60\%\sim85\%$RH , it is much higher than RH sensing using a bare silicon PC cavity, which is measured as 7$\times$10$^{-4}$ nm/\%RH. 
%To the best of our knowledge, by considering the RH-induced wavelength variations, this is the best sensitivity of optical RH sensors based on optical cavity and fiber Bragg grating. 

\vspace{12pt}

The RH-dependent sensing performance could be explained by the permeation mechanism of water molecules into a GO flake~\cite{NairR2012}. For the few-layer GO flake, relatively large spacings are kept by the groups  (hydroxyl, carbonyl, etc.) attached to graphene sheets, which also leave percolating regions of the nonoxidized graphene sheets. These interlayer spacings form a network of flow-channels for water molecules  moving with low-friction. The humidity could change the distance of GO's interlayer spacings. For instance, the distance is changed from 0.7 nm to 1.1 nm for RH varying from 30\% to 100\%~\cite{Cerveny2010,Lerf2006}. To allow a monolayer water molecule to permeate into GO's interlayer channel, a distance larger than 0.6 nm is required. As indicated in Ref.~\cite{NairR2012}, the distance variations with respect to the RH levels is not a linear function, presenting an abrupt rise around 60\%RH. For the GO-PC cavity, as the RH increases, more water molecules permeate into the GO layers for the gradually enlarged interlayer distance, which results in an increased effective refractive index of the GO flake. According to the electromagnetic perturbation theory of a resonant mode~\cite{Joannopoulos08}, the positive variation of the dielectric function in GO flake yields red-shifted resonant wavelength of the PC cavity, as observed in our experiments. Also, the RH-dependent wavelength variation slopes are consistent with the interlayer distances determined by the RH levels, as demonstrated in Ref.~\cite{NairR2012}. 

The stability and repeatability of the  sensing performance are examined by repeated measurements more than 100 times during a period of one month. The GO-PC cavity is kept in an ambient environment after each measurement. For each test, the humidity is changed between 20\% and 85\% RH with ascending and descending orders. As shown in Fig.~\ref{fig:spe}(d), the performance is stable over repeated measurements with small standard errors and small hysteresis at each RH reading. The GO-PC cavity could be reversible completely between low and high RH levels, and does not show degradation with time. It can be attributed to the superpermeability of the GO flake to water molecules and its nanoscale thickness.
%The above measurements are implemented with the chamber temperature stabilized at 25 $^\circ$C. For the realistic applications of a RH sensor, environment temperature variation is unavoided, which results in resonant wavelength shifts of the silicon PC cavity as well and introduces a measurement error. To estimate that, we measure the resonant wavelength shift of the PC cavity for different temperature, which exhibits a value of 0.036 nm/$^\circ$C. 

 \begin{figure}[th!]\centering
\includegraphics[width=6in]{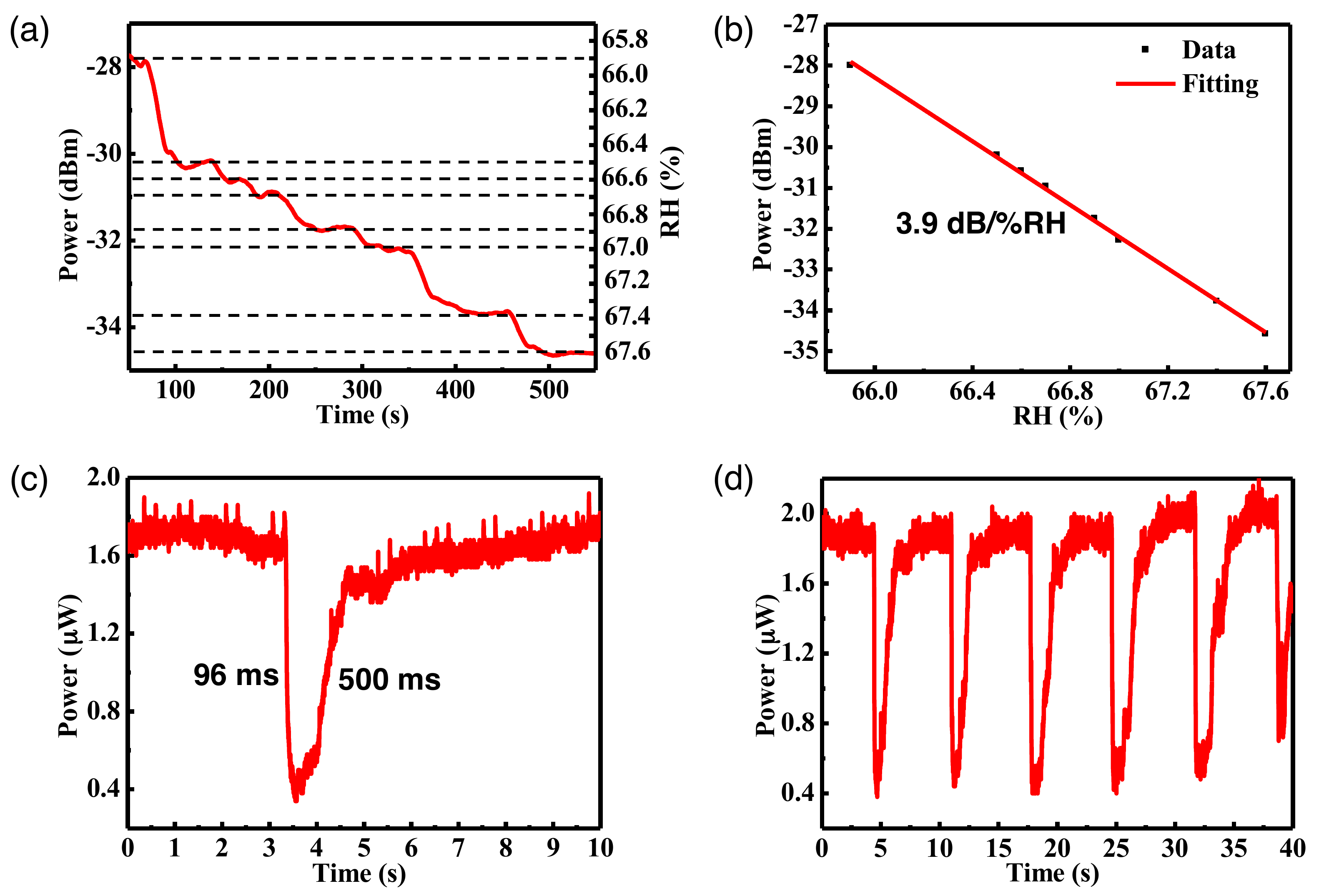}
\caption{{\small (Color online) (a) Measured powers of the cavity reflection when the RH is varied gradually. (b) RH-dependent power variations, showing a sensitivity of 3.9 dB/\%RH. (c) Response and recovery times of the sensor. (d) Periodic power variations of the sensor during the repeated human breathing, showing a complete recovery and repeatability. }}
\label{fig:pow} 
\end{figure}
\vspace{12pt}

For an optical RH sensor, implementations with an interrogation of optical power variations without wavelength-demodulation is also desired, though the wavelength interrogation could provide more reliable sensings. Because the resonant peak of the GO-PC cavity has a narrow linewidth, the wavelength shift induced by the varied RH could yield a remarkable variation of the reflection power when the excitation laser is on-resonance or slight off-resonance. To elucidate that, we input a laser at the wavelength of 1517.2 nm, and monitor the reflection power by changing the humidity around 67\%RH, as shown in Figs.~\ref{fig:pow}(a) and~\ref{fig:pow}(b). We observe the power changes even for a minor RH variation of 0.1\%, which is the limited RH resolution that our commercial RH sensor could calibrate. Around the resonant peak, the power variation versus RH represents a linear function with a slope of 3.9 dB$/\%$RH.

%, which has the best performance in the optical RH sensor interrogated by optical power. 

Because of the abundant hydrophilic functional groups on GO sheet and large interlayer spacing for low-friciton water flow, the GO-PC cavity sensor is expected to operate with fast response. To evaluate that, we monitor the cavity reflection power by applying a human breathing directly over the device from a distance further than 30 cm, and the laser wavelength is 1517.2 nm. As shown in Fig.~\ref{fig:pow}(c), a sharp power-decrease happens with the humidity breathing air interacts with the GO-PC cavity. The response time is estimated as 96 ms between the 10\% and 90\% power variations. After the breathing stops, the optical power recovers in a time interval of $\sim$500 ms. The longer recovery time could be attributed to the stronger interaction between the water molecules and GO layers than that of dry air and GO layers. Regarding the microscale dimension of the GO flake coated on the PC cavity, shown in Fig.~\ref{fig:sche}(d), the fast response time is consistent with the water flow velocity of 0.01 cm/s estimated in Ref.~\cite{NairR2012}. In Fig.~\ref{fig:pow}(d), we plot the monitored power variations during the several periods of human breathing. The periodically reversible maximum and minimum powers indicate the repeatability and reliability of the GO-PC cavity RH sensor. 

In conclusion, an optical humidity sensor is presented here based on a GO flake coated PC cavity. High sensing performances are facilitated by the strong light-GO interaction and microscale device dimension. The resonant wavelength of the GO-PC cavity is significantly shifted under high RH level due to the enlarged interlayer spacings of the GO flake. A high wavelength-shift rate of 0.68 nm/\%RH is obtained in the range of 60\% to 85\%RH. Relying on the resonant peak's narrow linewidth, the power interrogation of the sensor also presents a high sensitivity of 3.9 dB/$\%$RH. The response time is estimated less than 100 ms for the small device footprint, which is fast and sensitive enough to measure human breathing successfully. The good reliability and repeatability are examined as well. In the present geometry, the optical signal is vertically coupled through the far-field pattern of the PC cavity. We note an alternative integrated signal-processing device could be realized by coupling the cavity signal through a side-coupled PC waveguide~\cite{Noda2003Nature}. The GO-assisted humidity-activity could also be integrated onto a PC waveguide, which provides slow-light effect and sharp transmission bandedge to faciliate a high-performance RH sensor~\cite{Casas-Bedoya}.  Combining with the low-cost and easy integration of GO flakes, COMS-compatible fabrication of silicon PC cavity, and all-optical operation, the proposed GO-PC cavity RH sensor have potentials in situations where device compactness is required and the humidity changes dynamically, such as on-chip moisture monitoring and various medical diagnostics, including monitoring human exhaled breath.

Acknowledgment: Financial support was provided by the NSFC (61522507, 11404264, 61377035), and the Natural Science Basic Research Plan in Shaanxi Province of China (2016JQ6004).

%\begin{thebibliography}{99}
%\bibitem{Sander} D. Sander, M. O. Dücker, O. Blume, and J. Muller, Proc. SPIE {\bf 2686}, 100 (1996). 
%\end{thebibliography}

%\bibliographystyle{unsrt}

%\bibliography{library.bib}

\end{document}